\documentclass[letterpaper]{article} 
\usepackage{nips14submit_e,times}
\usepackage{url}

\usepackage{mysetup}

\title{Classification and Bayesian Optimization for Likelihood-Free Inference}

\author{
Michael U. Gutmann and Jukka Corander\\
Dept of Mathematics and Statistics\\
University of Helsinki and HIIT, Finland\\
\texttt{\{michael.gutmann, jukka.corander\}@helsinki.fi} \\
\And
Ritabrata Dutta and Samuel Kaski\\
 Dept of Information and Computer Science \\
 Aalto University and HIIT, Finland\\
\texttt{\{ritabrata.dutta, samuel.kaski\}@aalto.fi} \\
}

\nipsfinalcopy 

\begin{document}

\maketitle

\begin{abstract}
  Some statistical models are specified via a data generating process
  for which the likelihood function cannot be computed in closed
  form. Standard likelihood-based inference is then not feasible but
  the model parameters can be inferred by finding the values which
  yield simulated data that resemble the observed data. This approach
  faces at least two major difficulties: The first difficulty is the
  choice of the discrepancy measure which is used to judge whether the
  simulated data resemble the observed data. The second difficulty is
  the computationally efficient identification of regions in the
  parameter space where the discrepancy is low. We give here an
  introduction to our recent work where we tackle the two difficulties
  through classification and Bayesian optimization.
\end{abstract}

\section{Introduction}
The likelihood function plays a central role in statistics and machine
learning. It is the joint probability of the observed data $\Dobs$
seen as
a function of the model parameters of interest $\mytheta$. We may
assume that the data $\Dobs$ are a realization of a two stage sampling
process,
\begin{align}
\Z &\sim p_1(\Z | \thetaTrue),&  \Dobs &\sim p_2(\Dobs | \Z, \thetaTrue),
\end{align}
where $\Z$ are unobserved variables and $\thetaTrue$ are some fixed
but unknown values of the model parameters. The likelihood function
$L(\mytheta)$ is implicitly defined via an integral,
\begin{equation}
  L(\mytheta) = p(\Dobs | \mytheta) = \int p_2( \Dobs |
  \Z, \mytheta) p_1(\Z| \mytheta) \ud \Z.
\end{equation}
For many realistic data generating processes, the integral cannot be
computed analytically in closed form, and numerical approximation is
computationally too costly as well. Standard likelihood-based
inference is then not feasible. But inference can be performed by
using the possibility to simulate data from the model. Such
simulation-based likelihood-free inference methods have emerged in
multiple disciplines: ``Indirect inference'' originated in economics
\citep{Gourieroux1993}, ``approximate Bayesian computation'' (ABC) in
genetics \citep{Beaumont2002, Marjoram2003, Sisson2007}, or the
``synthetic likelihood'' approach in ecology \citep{Wood2010}. The
different methods share the basic idea to identify the model
parameters by finding values which yield simulated data that resemble
the observed data. The inference process is shown in a schematic way
in Algorithm \ref{alg:ABC} in the framework of ABC.

In Algorithm \ref{alg:ABC}, two fundamental difficulties of the
aforementioned inference methods are highlighted. One difficulty is
the measurement of similarity, or discrepancy, between the observed
data $\Dobs$ and the simulated data $\Dsimtheta$ (line 5). The choice
of discrepancy measure affects the statistical quality of the
inference process. The second difficulty is of computational
nature. Since simulating data $\Dsimtheta$ can be computationally very
costly, one would like to identify the region in the parameter space
where the simulated data resemble the observed data as quickly as
possible, without proposing parameters $\mytheta$ which have a
negligible chance to be accepted (line 3).

We have been working on both problems, the choice of the discrepancy
measure and the fast identification of the parameter regions of
interest \citep{Gutmann2014b,Gutmann2015a}. The following two sections
are a brief introduction to the two papers.
\begin{algorithm}[t]
\SetKwComment{Comment}{}{}
  \DontPrintSemicolon
  \KwResult{$N$ samples $\boldsymbol{\theta}_1 \ldots \boldsymbol{\theta}_N$}
  \For{i = 1 to N}{ \vspace{1ex}
    \Repeat{ \red{$\Dobs \approx \Dsimtheta$}\Comment*[f]{{\small \it \red{affects the quality of the inference}}} }{
      \red{Propose parameter $\mytheta$} \Comment*{{\small \it \red{affects the speed of the inference}}}
      Generate pseudo observed data $\Dsimtheta \sim p(\Dsimtheta| \mytheta)$\;
    }
    \vspace{1ex}
  set $\boldsymbol{\theta}_i = \boldsymbol{\theta}$\;
 }
 \caption{\label{alg:ABC} Basic ABC algorithm.}
\end{algorithm}

\section{Discriminability as discrepancy measure}
We transformed the original problem of measuring the discrepancy
between $\Dsimtheta$ and $\Dobs$ into a problem of classifying the
data into simulated versus observed \citep{Gutmann2014b}. Intuitively,
it is easier to discriminate between two data sets which are very
different than between data which are similar, and when the two data
sets are generated with the same parameter values, the classification
task cannot be solved significantly above chance-level. This motivated
us to use the discriminability (classifiability) as discrepancy
measure, and to perform likelihood-free inference by identifying the
parameter values which yield chance-level discriminability only
\citep{Gutmann2014b}. 

\begin{figure}[t]
\subfloat[Gaussian densities]{\includegraphics[width = 0.48\textwidth]{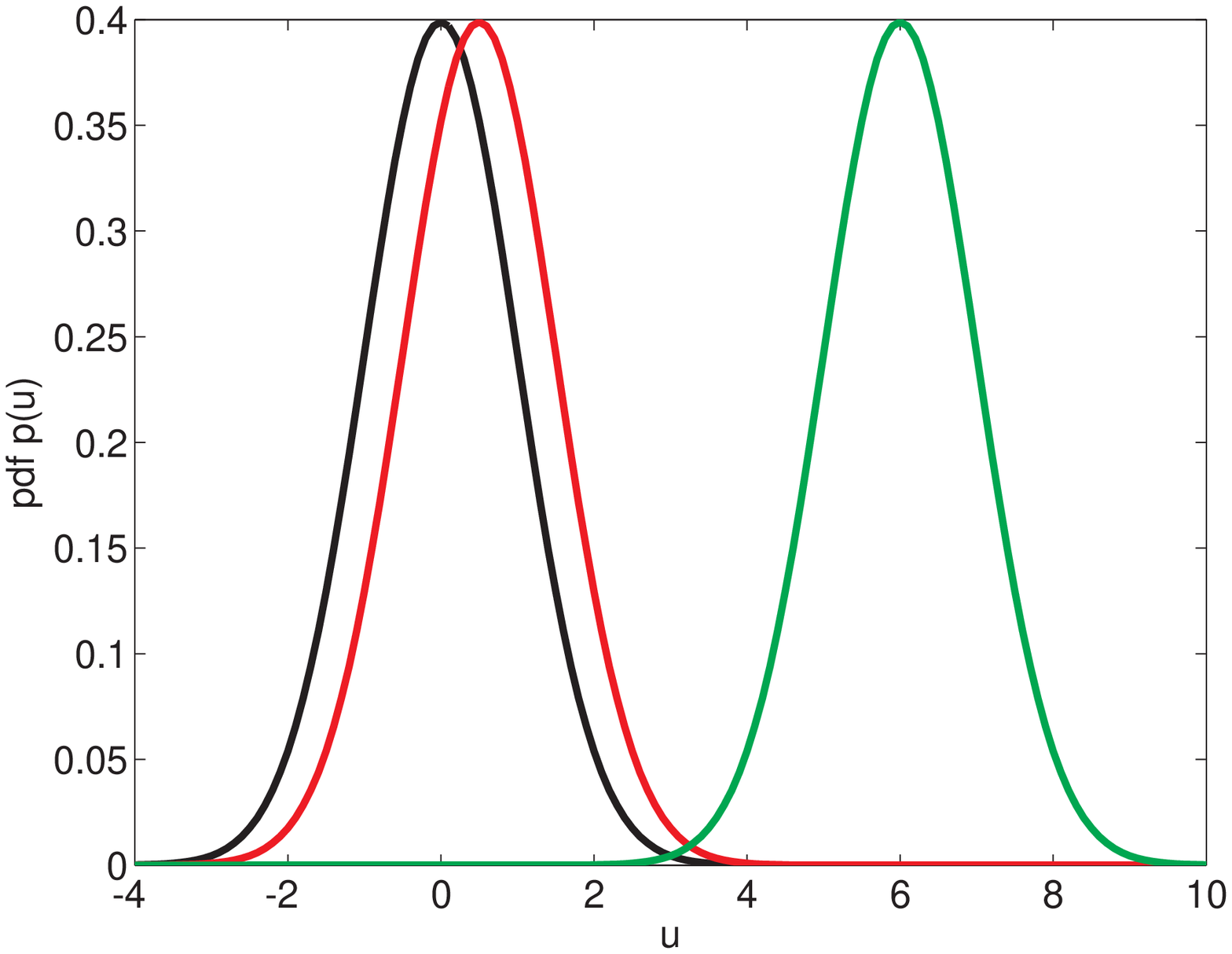}}\hfill
\subfloat[Classification performance of LDA]{\includegraphics[width = 0.48\textwidth]{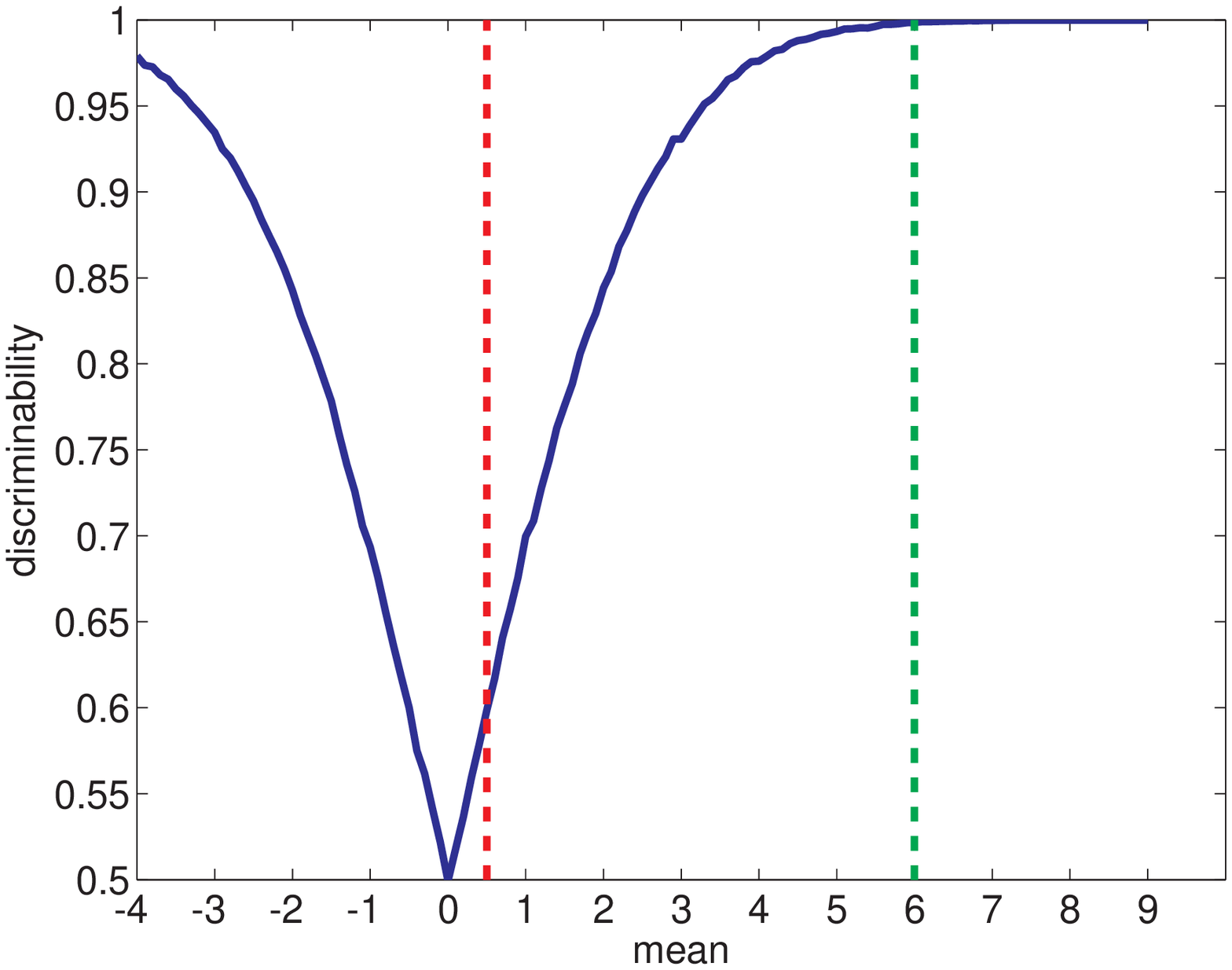}}
\caption{\label{fig:Gauss_large_sample} Discriminability as discrepancy measure, illustration on toy data  ($n=\np{10000}$).}
\end{figure}

We next illustrate this approach using a toy example. The data $\Dobs = (x_1,\ldots,x_n)$
are assumed to be sampled from a standard normal distribution (black
curve in Figure \ref{fig:Gauss_large_sample}(a)), and the parameter of
interest $\theta$ is the mean. For data simulated with mean $\theta=6$
(green curve), the two densities barely overlap so that classification
is easy. In fact, linear discriminant analysis (LDA) yields a
discriminability of almost 100\% (Figure
\ref{fig:Gauss_large_sample}(b), green dashed curve). If the data are
simulated with a mean closer to zero, for example with $\theta=1/2$
(red curve), the simulated data $\Y_\theta$ become more similar to
$\Dobs$ and the classification accuracy drops to around 60\% (red dashed
curve). For $\theta=0$, where the simulated and observed data are
generated with the same values of $\theta$, only chance-level
discriminability of 50\% is obtained. This illustrates how
discriminability can be used as a discrepancy measure.

We analyzed the validity of this approach theoretically and
demonstrated it on more challenging synthetic data as well as real
data with an individual-based epidemic model for bacterial infections
in day care centers \citep{Gutmann2014b}. The finding that
classification can be used to measure the discrepancy has both
practical and theoretical value: The main practical value is that the
rather difficult problem of choosing a discrepancy measure is reduced
to a more standard problem where we can leverage on effective
existing solutions. The theoretical value lies in the
establishment of a tight connection between likelihood-free inference
and classification -- two fields of research which appear rather different at first glance.

\section{Bayesian optimization to identify parameter regions of interest}
\begin{figure}[t]
\centering
\includegraphics[width = 0.48\textwidth]{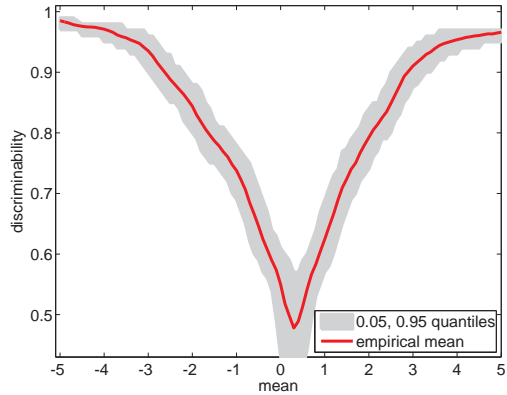}
\caption{\label{fig:random_proc} Distribution of $\Deltatheta$ for the
  Gaussian toy example using $n=50$.}
\end{figure}

In the following, we denote a certain discrepancy measure by
$\Deltatheta$. A small value of $\Deltatheta$ is assumed to imply that
$\Dsimtheta$ are judged to be similar to $\Dobs$. The difficulty in
finding parameter regions where $\Deltatheta$ is small is at least
twofold: First, the mapping from $\mytheta$ to $\Deltatheta$ can
generally not be expressed in closed form and derivatives are not
available either. Second, $\Deltatheta$ is actually a stochastic
process due to the use of simulations to obtain $\Dsimtheta$. We
illustrate this in Figure \ref{fig:random_proc} for our
Gaussian toy example where $\Deltatheta$ is the discriminability
between $\Dobs$ and $\Dsimtheta$  \citep[for further examples, see][]{Gutmann2015a}. The figure visualizes the
distribution of $\Deltatheta$ for $n=50$. The fact that $\Deltatheta$
is a random process was suppressed in Figure
\ref{fig:Gauss_large_sample} by working with a large sample size.

We used Bayesian optimization, a combination of nonlinear (Gaussian
process) regression and optimization \citep[see, for example,
][]{Brochu2010}, to quickly identify regions where $\Deltatheta$ is
likely to be small \citep{Gutmann2015a}. In Bayesian optimization, the
available information $\{(\mytheta^{(k)},\Deltatheta^{(k)}),
k=1,\ldots,K\}$ about the relation between $\mytheta$ and
$\Deltatheta$ is used to build a statistical model of $\Deltatheta$,
and new data are actively acquired in regions where the minimum of
$\Deltatheta$ is potentially located. After acquisition of the new
data, e.g. a tuple $(\mytheta^{(K+1)},\Deltatheta^{(K+1)})$, the model
is updated using Bayes' theorem.

\begin{figure}[t]
\centering
\subfloat[Model after one acquisition]{\includegraphics[width = 0.49\textwidth]{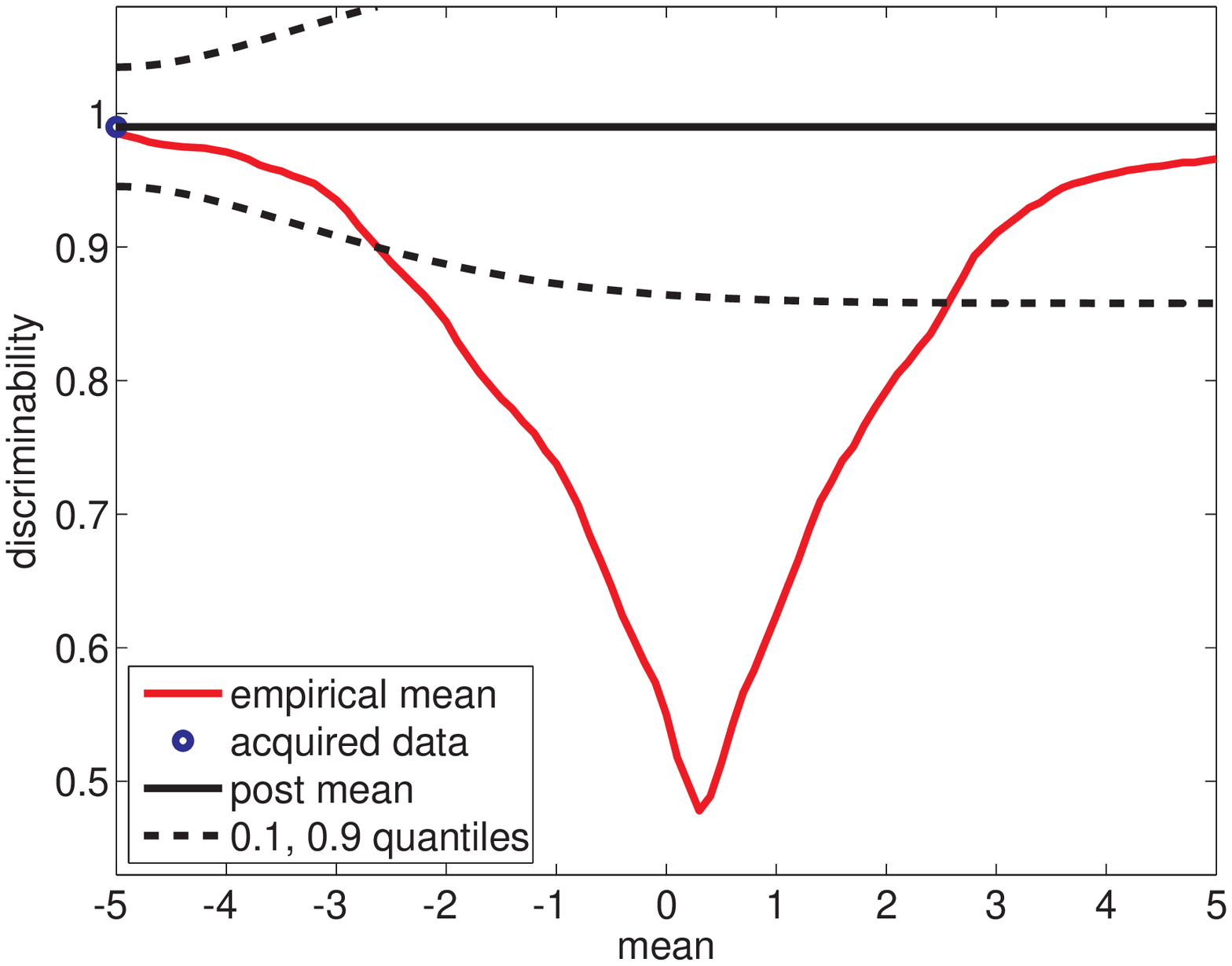}}\hfill
\subfloat[Model after two acquisitions]{\includegraphics[width = 0.49\textwidth]{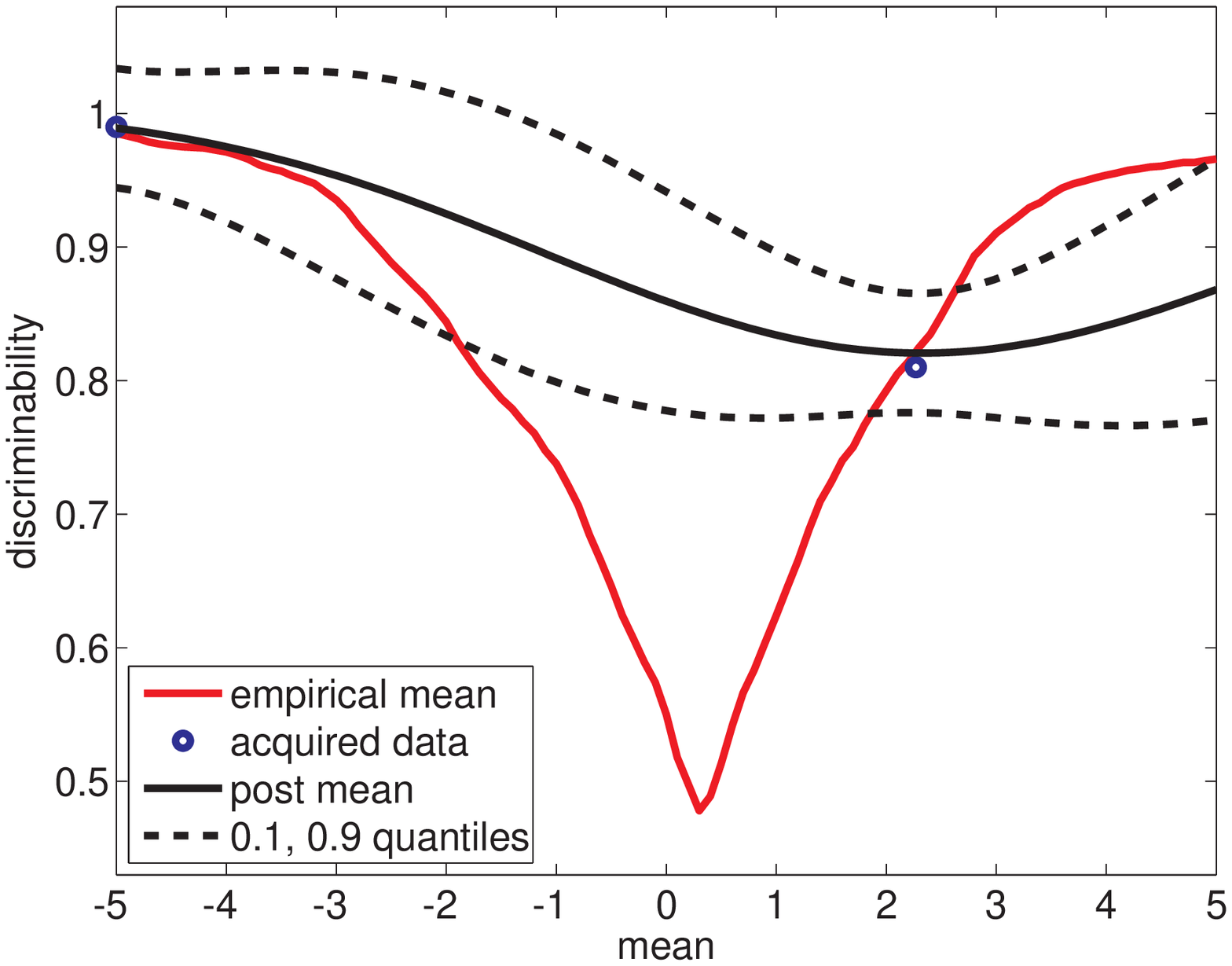}}\\
\subfloat[Model after four acquisitions]{\includegraphics[width = 0.49\textwidth]{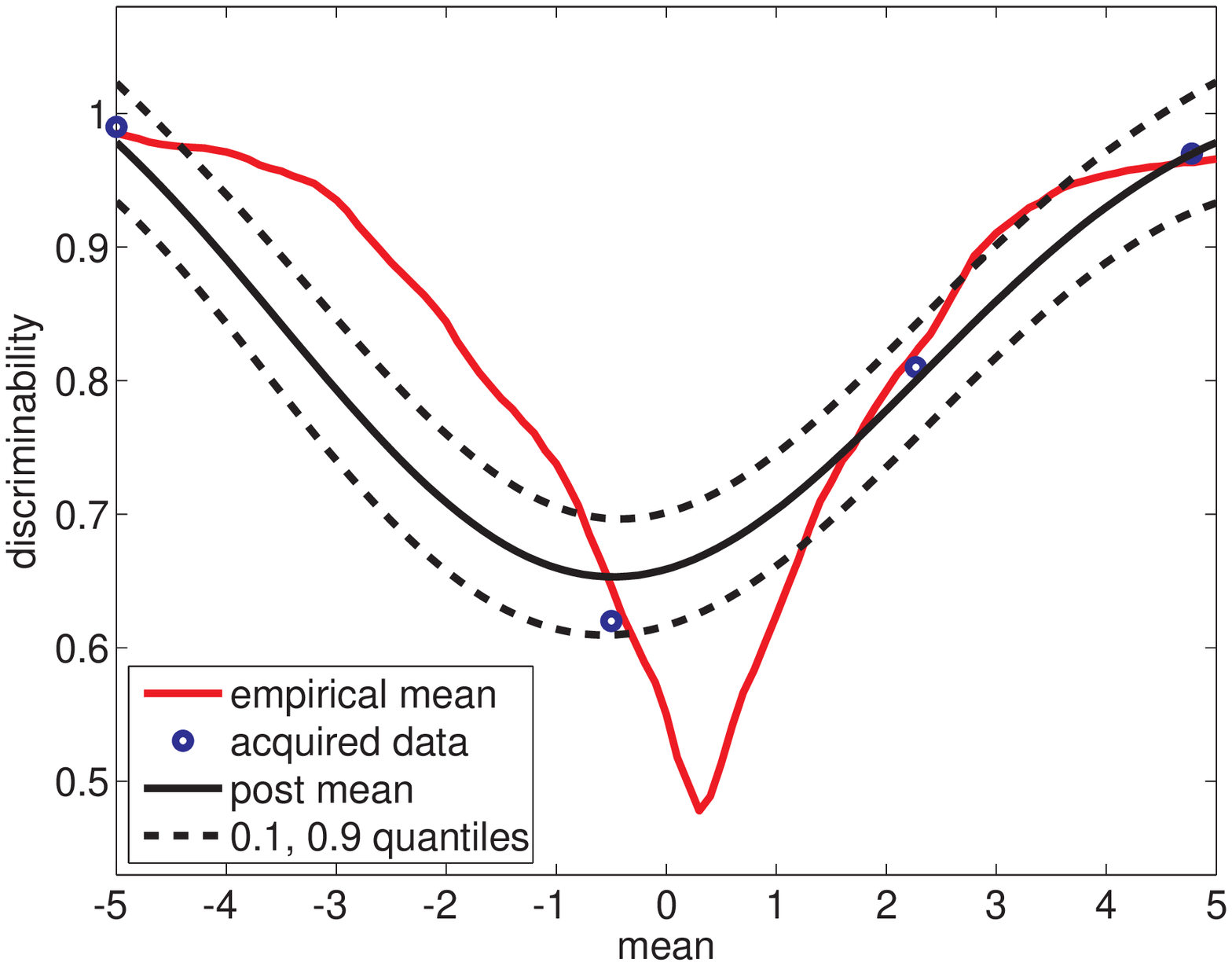}}\hfill
\subfloat[Model after eight acquisitions]{\includegraphics[width = 0.49\textwidth]{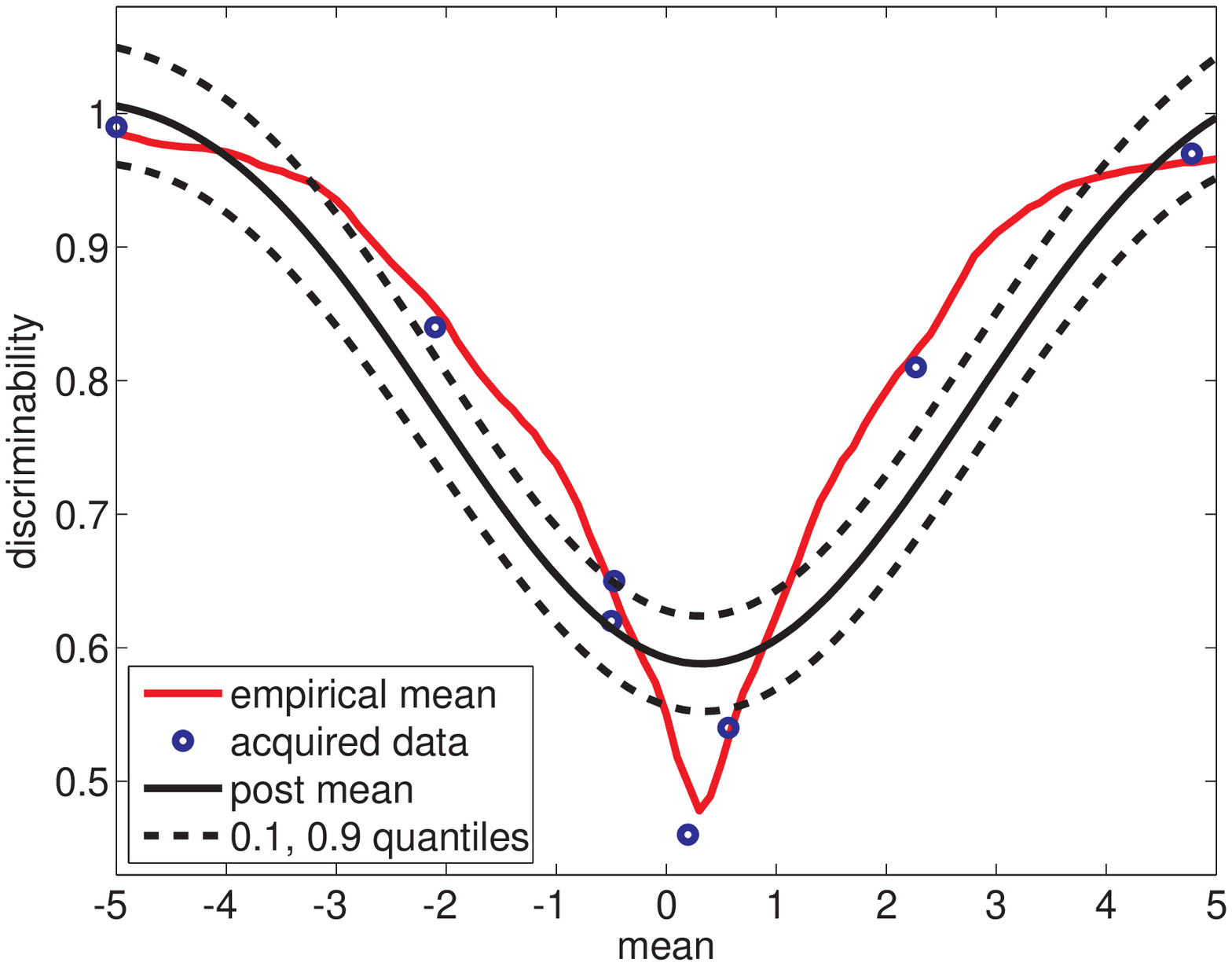}}\\
\subfloat[Model after ten acquisitions]{\includegraphics[width = 0.49\textwidth]{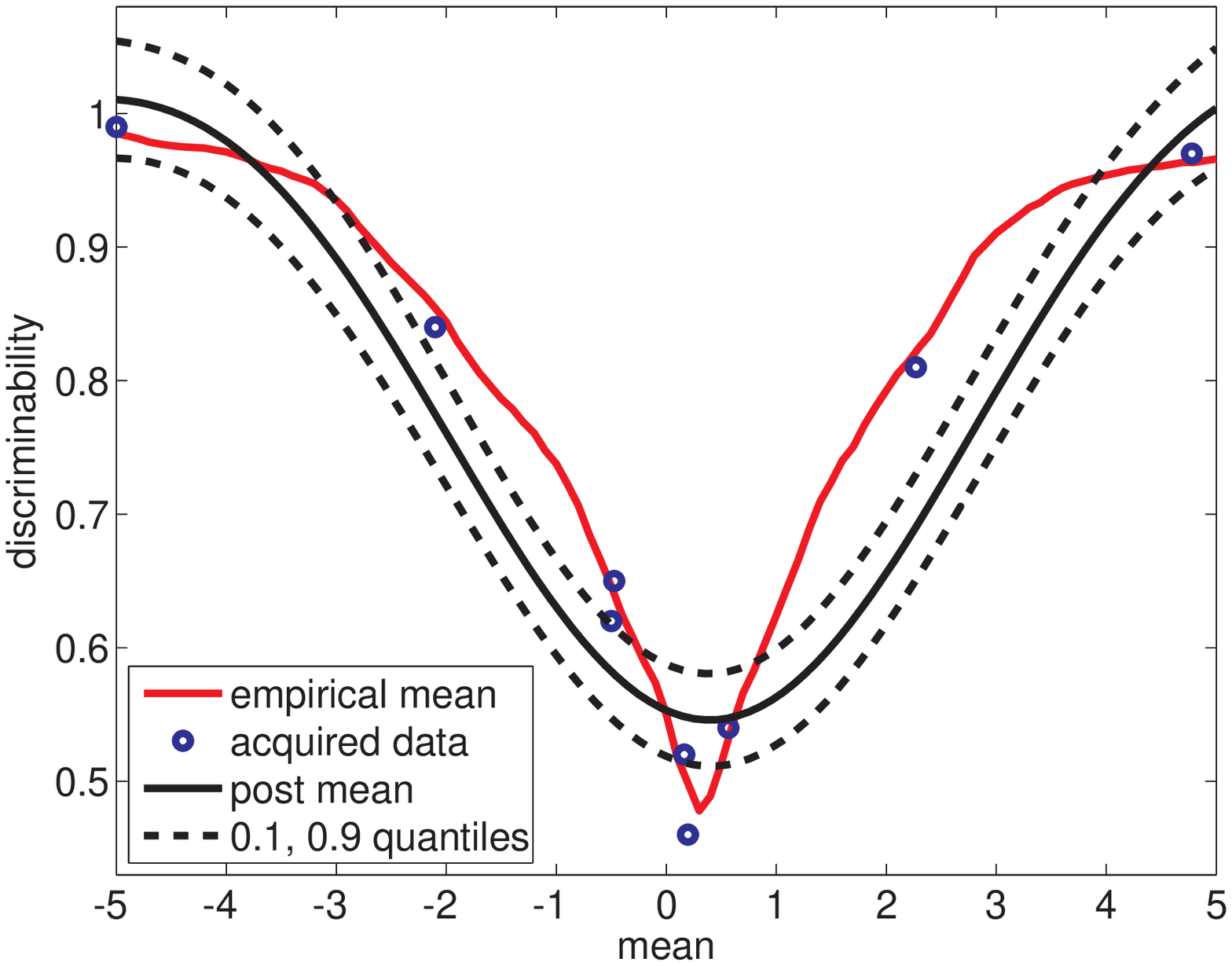}}\hfill
\subfloat[Model after twenty acquisitions]{\includegraphics[width = 0.49\textwidth]{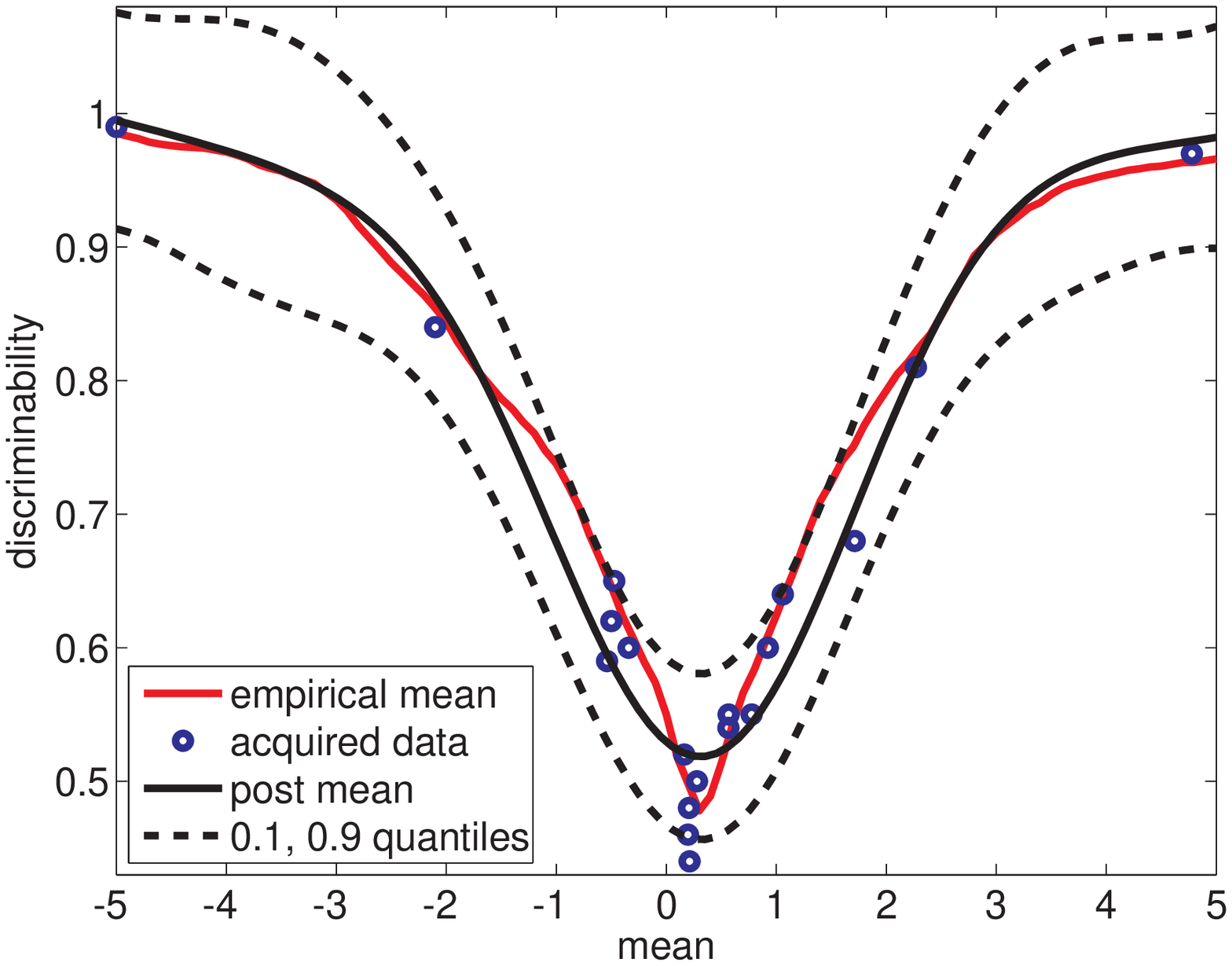}}
\caption{\label{fig:Gauss_small_sample}Bayesian optimization to
  quickly identify the parameter regions of interest.}
\end{figure}

For our simple toy example, the region around zero was identified as
the region of interest within ten acquisitions (Figure
\ref{fig:Gauss_small_sample}(a-e)). While the location of the minimum is
approximately correct, the posterior mean approximates the (empirical)
mean of $\Deltatheta$ in Figure \ref{fig:random_proc} only roughly. As more evidence
about the behavior of $\Deltatheta$ in the region of interest is
acquired, the fit improves (Figure \ref{fig:Gauss_small_sample}(f)).

In the full paper \citep{Gutmann2015a}, we show that Bayesian
optimization not only allows to quickly identify the regions of interest
but also to perform approximate posterior inference. Our findings are
supported by theory, and applications to real data analysis with
intractable models. In our applications, the inference was accelerated
through a reduction in the number of required simulations by several
orders of magnitude.
\section{Conclusions}
Two major difficulties in likelihood-free inference are the choice of
the discrepancy measure between simulated and observed data, and the
identification of regions in the parameter space where the discrepancy
is likely to be small. The former difficulty is more of statistical,
the latter more of computational nature. We gave a brief introduction
to our recent work on the two issues: We used classification to
measure the discrepancy \citep{Gutmann2014b}, and Bayesian
optimization to quickly identify regions of low discrepancy
\citep{Gutmann2015a}.
\bibliographystyle{plainnat}
\bibliography{refs}

\end{document}